# Characterization and manipulation of individual defects in insulating hexagonal boron nitride using scanning tunneling microscopy

Dillon Wong[1]*, Jairo Velasco Jr.[1]*, Long Ju[1]*, Juwon Lee[1], Salman Kahn[1], Hsin-Zon Tsai[1], Chad Germany[1], Takashi Taniguchi[4], Kenji Watanabe[4], Alex Zettl[1,2,3], Feng Wang[1,2,3] and Michael F. Crommie[1,2,3]

[1]*Department of Physics, University of California, Berkeley, California 94720, USA*
[2]*Materials Sciences Division, Lawrence Berkeley National Laboratory, Berkeley, California 94720, USA*
[3]*Kavli Energy NanoSciences Institute at the University of California, Berkeley and the Lawrence Berkeley National Laboratory, Berkeley, California 94720, USA*
[4]*National Institute for Materials Science, 1-1 Namiki, Tsukuba, 305-0044, Japan*

*These authors contribute equally to this manuscript.
[†] *Email:crommie@berkeley.edu*

**Defects play a key role in determining the properties of most materials and, because they tend to be highly localized, characterizing them at the single-defect level is particularly important. Scanning tunneling microscopy (STM) has a history of imaging the electronic structure of individual point defects in conductors[1], semiconductors[2-7], and ultrathin films[8], but single-defect electronic characterization at the nanometer-scale remains an elusive goal for intrinsic bulk insulators. Here we report the characterization and manipulation of individual native defects in an intrinsic bulk hexagonal boron nitride (BN) insulator via STM. Normally, this would be impossible due to the lack of a conducting drain path for electrical current. We overcome this problem by employing a graphene/BN heterostructure, which exploits graphene's atomically thin nature to allow visualization of defect phenomena in the underlying bulk BN. We observe three different defect structures that we attribute to defects within the bulk insulating boron nitride. Using scanning**

**tunneling spectroscopy (STS), we obtain charge and energy-level information for these BN defect structures. In addition to characterizing such defects, we find that it is also possible to manipulate them through voltage pulses applied to our STM tip.**

BN is an essential component in many new devices[9-11] that incorporate two-dimensional materials. Therefore it is crucial to understand the nature of intrinsic defects in BN. Previous cathodoluminescence and elemental analysis of BN indicated the existence of residual impurities and defects in high-purity single crystal BN synthesized at high pressure and temperature[11,12]. Optoelectronic experiments have revealed that these defects give rise to photoactive states within the BN band gap[13,14]. So far, however, these studies have been limited to spatially averaged defect behavior, and investigation of individual defects at the nanoscale remains an outstanding challenge.

Here we visualize individual BN defects by capping a BN crystal with a monolayer of graphene. Fig. 1a shows a typical STM topographic image of our graphene/BN heterostructures, where a 7 nm moiré pattern can be seen on top of long-range height fluctuations spanning tens of nanometers, similar to previous imaging of graphene on BN[15,16]. Localized shallow dips and a protrusion are also visible ($\Delta z < 0.1$ Å). More revealing, however, are the differential conductance (d$I$/d$V$) maps shown in Fig. 1b. Striking new features are visible in these data. We observe randomly distributed bright (high d$I$/d$V$) and dark (low d$I$/d$V$) circular dots of ~20 nm in diameter that have varying degrees of intensity (see Supplement for densities and intensities). Another common feature, as seen in the right edge of the map in Fig. 1b, is a sharp ring structure with an interior that does not obscure the moiré pattern. Close-up topographic studies of these defects reveal unblemished atomically resolved graphene honeycomb structure with occasional slight dips or a protrusion having $|\Delta z| < 0.1$ Å (see Supplement). Maps obtained at numerous locations with many tips across three different devices replicate these observations.

Figs 2a and 2b show higher resolution d$I$/d$V$ maps of representative bright and dark dot defects. These maps show clearly that the graphene moiré pattern is not obscured by the defects. To determine the effect of these defects on the electronic structure of graphene, we performed d$I$/d$V$ spectroscopy at varying distances from the dot centers (each spectrum was started with the same tunnel current $I$ and sample bias $V_s$). These data are plotted in Figs 2c and 2d for the bright and dark dots, respectively. The spectra are characteristic of undamaged graphene[17], but show an electron/hole asymmetry that is dependent on the tip position relative to the center of a defect. In Fig. 2c, for example, we see that d$I$/d$V$ ($V_s > 0$) increases as the tip approaches the bright-dot center. Fig. 2d shows the opposite trend as seen by the decrease in d$I$/d$V$ ($V_s > 0$) as the STM tip approaches the dark-dot center. These basic trends were seen for all bright and dark dot defects, regardless of intensity and tip-height configuration (see Supplement). These observations can be understood by recalling that d$I$/d$V$ reflects the graphene local density of states (LDOS). The distance-dependent enhancement of d$I$/d$V$ above the Dirac point ($V_s \approx -0.17$ V) as the tip nears a bright dot in Fig. 2c can therefore be interpreted as arising from the attraction of negatively charged Dirac fermions to the dot center. We thus conclude that the bright dot in Fig. 2a reflects a positively charged defect in BN[18,19]. Similarly, the distance-dependent reduction of d$I$/d$V$ above the Dirac point in Fig. 2d arises from the repulsion of negatively charged Dirac fermions from the defect. We thus conclude that the dark dots are negatively charged[18,19].

We now focus on the ring defects, as displayed at the right edge of Fig. 1b. We find that the ring radius depends on the values of $V_s$ and back-gate $V_g$. Fig. 3 shows that the ring radius changes from 2 nm (Fig. 3a) to 11 nm (Fig. 3b) as $V_g$ is changed from $V_g = 17$ V to $V_g = 9$ V (for a constant $V_s = -0.3$ V). Fig. 3c shows the dependence of the ring radius on $V_g$ for various $V_s$ values (denoted by distinct symbols). These data were obtained by measuring the ring radius from d$I$/d$V$ maps taken at the same location as Figs 3a and 3b, but with different $V_s$ and $V_g$ configurations. Although the precise ring radius depends on the sharpness of the STM tip[4], the qualitative behavior shown in Fig. 3c is typical of the vast majority of ring defects observed here. In general, for fixed $V_s$, the ring radius increases with decreasing $V_g$ until a critical back-gate voltage ($V_c = 6 \pm 1$V) is reached, upon which the ring vanishes.

We now discuss the origin of the "dot" and "ring" defects observed in our d$I$/d$V$ maps. Three general scenarios are possible: (i) Adsorbates bound to the surface of graphene, (ii) adsorbates trapped at the interface between graphene and BN, and (iii) intrinsic defects within the insulating BN substrate. Our data imply that (iii) is the correct scenario, as follows: First we rule out scenario (i) because weakly bound adsorbates would have a higher height profile than the topographically small features observed[18,20] (see Supplement), and would also likely get swept away by the STM tip when it is brought close enough to observe the graphene honeycomb structure[19]. Strongly bound adsorbates in scenario (i) would also likely have taller height profiles as seen for other graphene adsorbates (see Supplement) and should disrupt the graphene honeycomb lattice[21,22] (which was not observed). Also, strongly bound adsorbates should lead to changes in the graphene spectroscopy due to formation of localized bonding states[22], which are not seen. Scenario (ii) can be ruled out because an adsorbate trapped beneath graphene would cause a bump in graphene at least an order of magnitude larger than the $\Delta z < 0.1$ Å feature observed here[23]. We would also expect a trapped adsorbate to locally delaminate the graphene from the BN substrate, thus disrupting the moiré pattern, which is not seen.

Scenario (iii), intrinsic charged BN defects, is thus the most likely explanation of the defects observed here. Polycrystalline BN has been shown to host several varieties of charged defects, as seen from electron paramagnetic resonance [24] and luminescence experiments[25,26], as well as theoretical investigations[27]. In those studies the most abundantly reported defects were N-vacancies, which were shown to act as donors, and C impurities substituted at N sites, which were shown to act as acceptors. Secondary ion mass spectroscopy studies of high-purity single crystal BN synthesized at high pressure and temperature have identified oxygen and carbon impurities[12]. Comparison between optoelectronic experiments[13,14] of high-purity single crystal BN and recent theoretical work[28] shows that the nature of the BN crystal defects is consistent with the observations of C impurities and N-vacancies. Such defects, when ionized, could induce the bright and dark dots observed in graphene/BN via a graphene screening response[29] (Figs 1 and 2). The fact that these defects are embedded in the BN explains why the dots have such small topographic deflection, as well as why the graphene lattice and moiré pattern are not disrupted, and also why no new states arise in the graphene spectroscopy[20,22]. Variations in the intensity of bright and dark defects are explained by BN defects lying at different depths relative to the top graphene layer.

It is possible to extract quantitative information regarding the electronic configuration of BN defects from the STM d$I$/d$V$ signal measured from the graphene capping layer. This can be achieved for the ring defects by analyzing the gate ($V_g$) and bias ($V_s$) dependent ring radius, shown in Fig. 3c. Similar rings have been observed in other systems and have been attributed to the charging of an adsorbate or defect[4,20,30-32]. Because the ring in Fig. 3 is highly responsive to the presence of the STM tip and displays no charge hysteresis, we expect that it lies in the topmost BN layer and is strongly coupled to the graphene electronic structure. The STM tip is capacitively coupled to the graphene directly above the defect through the equation $|e|\delta n = C(r)V_{tip}$, where δn is the local change in graphene electron density, $C(r)$ is a capacitance that increases with decreasing lateral tip-defect distance $r$, and $V_{tip}$ is the tip electrostatic potential ($V_{tip} = -V_s$ + constant, see Supplement). For the d$I$/d$V$ maps in Fig. 3, $V_{tip} < 0$, so the electrostatic gating from the tip lowers the electron density of the (n-doped) graphene directly beneath the tip. Fig. 3d schematically depicts the local electronic structure of the graphene immediately above the defect when $r$ is large and $V_g$ is set such that the defect level is filled and carries negative charge. As the tip approaches the defect, $C(r)$ increases, and thus δn becomes more negative. Eventually the defect level crosses the Fermi level (and switches to a neutral state) when the tip is at a distance $R$ away from the defect, thus causing a perturbation in the tunnel current that leads to the observation of a ring of radius $R$. Fig. 3e shows the case ($r < R$) where the defect is in a neutral charge state through interaction with the tip. The energy level of the defect can be found by tuning $V_g$ such that the Fermi level matches the defect level in the absence of the tip. This will cause the radius of the charging ring to diverge. As seen in Fig. 3c, this occurs for the observed ring defects when $V_g = 6 \pm 1$ V, thus resulting in a defect level approximately $30 \pm 10$ meV above the graphene Dirac point energy (since the Dirac point energy can be measured with respect to the Fermi level), which is expected to be ~4 eV below the BN conduction band edge[33]. Interestingly, this is similar to a previously observed carbon substitution defect level[25], suggesting that the ring defect arises from a carbon impurity.

Additional microscopic information regarding the observed BN defects can be obtained by directly manipulating their charge state with the STM tip. Similar manipulation has been performed previously to switch the charge state of defects in semiconductors[34] as well as adatoms on top of ultrathin insulating films[8,35], but this type of STM-based manipulation is unprecedented for defects inside bulk insulators. Fig. 4a shows a d$I$/d$V$ map exhibiting numerous charged defects. In order to manipulate the charge state of the observed BN defects, the STM tip was positioned over the center point of this area and a bias of $V_s$ = 5 V was applied (see Supplement). After applying this voltage pulse, a d$I$/d$V$ map was acquired over the same region at low bias, as shown in Fig. 4b. Fig. 4c shows the same region after similar application of a second pulse. Inspection of Figs 4b and 4c shows that the BN defect configurations are significantly altered by application of such voltage pulses. The defects are seen to reversibly switch between charged and neutral states, as well as between states having opposite charge. To highlight this behavior, we denote changes to defect states (compared to the preceding image) with colored arrows. A red arrow signifies the disappearance of a charged defect, a blue arrow represents the appearance of a charged defect, and a green arrow indicates where a defect has changed the sign of its charge. We find that defects that disappear after a tip pulse always reappear in the same location after subsequent tip pulses. Additionally, dark dots tend to switch into metastable neutral states (i.e. disappear) at a higher rate than bright dots. Ring defects, as well as the darkest and brightest dots, remain unchanged by tip pulses.

This tip-induced manipulation of BN defects can be explained by electric-field-induced emission of charge carriers from BN defect states. By tilting the local potential landscape, the STM tip causes charge carriers to tunnel through the ionization barrier between different defects, charging some while neutralizing others (see Supplement). This accounts for the disappearance and reappearance of the dots in the same location, which cannot be described by defect migration through the BN lattice. In addition, the observation of a higher rate of switching for the dark dots (acceptors) than bright dots (donors) suggests it is more energetically favorable for neutral acceptors to emit holes than for neutral donors to emit electrons; hence the acceptor states are likely closer to the valence band than the donor states are to the conduction band. Since the rings, as well as the darkest and brightest dots, never change under tip pulses, we surmise that they are in the topmost layer of BN and in direct electrical contact with the graphene. Their charge states thus depend only on graphene's local chemical potential and show no hysteresis with electric field. The defects that switch into new metastable charge states (i.e., exhibit hysteresis) must thus be in lower BN layers, out of direct contact with the graphene.

In conclusion, we have shown that a single graphene capping layer enables STM imaging and manipulation of individual point defects in an insulating bulk BN substrate. We are able to identify the charge state of individual defects, and we have quantitatively extracted the energy level location for one species of defect (suggesting that it is a carbon impurity). We find that voltage pulses applied to our STM tip enable ionization, neutralization, and even toggling of the charge state for defects in BN. This new method of using an atomically thin conducting capping layer to probe and control defects in bulk insulators might be extended to other insulator/defect systems previously inaccessible to STM, such as diamond with nitrogen-vacancy centers.

**Methods:**

Our samples were fabricated with the transfer technique developed by Zomer and colleagues[36] using standard electron-beam lithography. We used BN crystals with 60-100 nm thickness and an $SiO_2$ thickness of 285 nm as the dielectric for electrostatic gating. Monolayer graphene was exfoliated from graphite and deposited onto methyl methacrylate (MMA) polymer and transferred onto BN (that was also annealed) sitting on an $SiO_2$/Si wafer. Completed devices were annealed in flowing Ar/$H_2$ forming gas at 350°C and their electrical conductance was measured with a standard ac voltage bias lock-in technique with a 50 μV signal at 97.13 Hz. Samples that exhibited good electron transport curves were then transferred into our Omicron ultra-high vacuum (UHV) low temperature STM. A second anneal was then performed for several hours at ~300°C and $10^{-11}$ torr before moving the device into the STM chamber for measurements at $T = 5$ K. Before all STM measurements, our platinum iridium STM tip was calibrated by measuring the surface state of an independent Au(111) crystal. All STM topographic images were acquired in a constant current mode with sample bias $V_s$ defined as the voltage applied to the sample with respect to the STM tip. All STS measurements were obtained by lock-in detection of the ac tunnel current induced by a modulated voltage (6 mV at 613 Hz) added to $V_s$.

**Acknowledgements:**

The authors thank P. Jarillo-Herrero, N. Gabor, A. Young, P. Yu and A. Rubio for stimulating discussions. This research was supported by the $sp^2$ program (STM measurement and instrumentation) and the LBNL Molecular Foundry (graphene characterization) funded by the Director, Office of Science, Office of Basic Energy Sciences of the US Department of Energy under Contract No. DE-AC02-05CH11231. Support also provided by National Science

Foundation award CMMI-1235361 (device fabrication, image analysis). J.V.J. acknowledges support from the UC President's Postdoctoral Fellowship. D.W. was supported by the Department of Defense (DoD) through the National Defense Science & Engineering Graduate Fellowship (NDSEG) Program. S.K. acknowledges support from the Qualcomm Scholars Research Fellowship.

**Author Contributions:**

L.J. and J.V.J. conceived the work and designed the research strategy. J.V.J., D.W., S.K. and J.L. performed data analysis. J.V.J., S.K., L.J. and A.Z. facilitated sample fabrication. D.W., J.L. and J.V.J. carried out STM/STS measurements. J.V.J. and S.K. carried out electron transport measurements. K.W. and T.T. synthesized the h-BN samples. D.W., J.V.J. and L.J. formulated theoretical model with advice from F.W. and M.F.C. M.F.C. supervised the STM/STS experiments. J.V.J., D.W. and M.F.C. co-wrote the manuscript. J.V.J. and M.F.C. coordinated the collaboration. All authors discussed the results and commented on the paper.

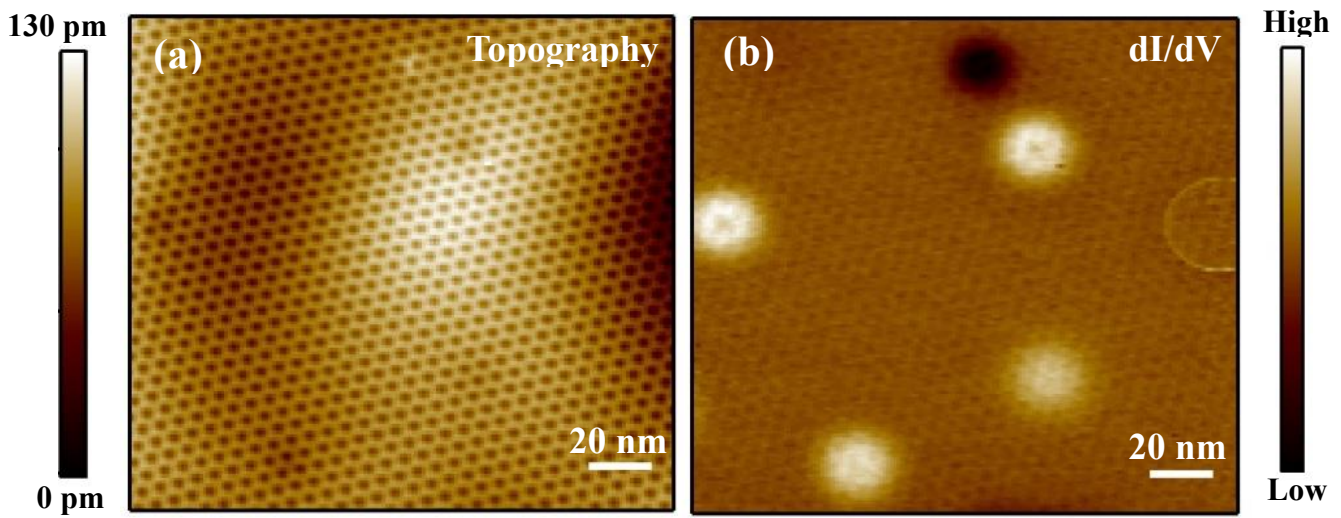

**Figure 1**: **STM topography and corresponding d$I$/d$V$ map for a graphene/BN device.** (a) STM topographic image of a clean graphene/BN area. (b) d$I$/d$V$ map ($I$ = 0.4 nA, $V_s$ = -0.25 V) acquired simultaneously with (a) exhibiting various new features: bright dots, a dark dot, and a ring.

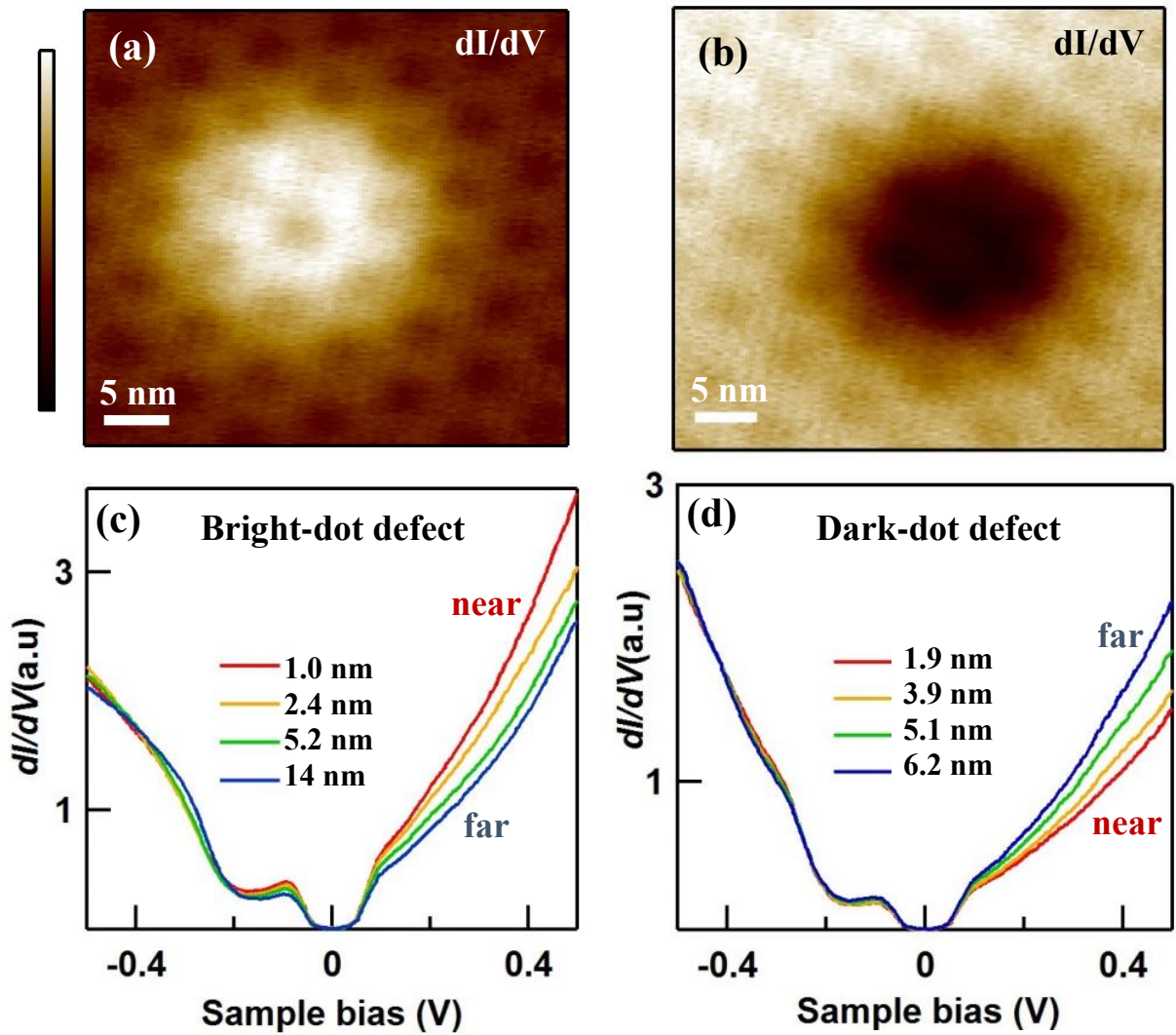

**Figure 2**: **d*I*/d*V* maps and spatially dependent d*I*/d*V* spectroscopy determines defect charge state.** (a-b) d*I*/d*V* maps ($I$ = 0.4 nA, $V_s$ = -0.3 V, $V_g$ = 5 V) for bright and dark dot defects. (c) d*I*/d*V* spectroscopy (initial tunneling parameters: $I$ = 0.4 nA, $V_s$ = -0.5 V, $V_g$ = 20 V) measured on graphene at different lateral distances from the center of the bright dot in (a). (d) Same as (c), but for the dark dot in (b). From the distance-dependent d*I*/d*V* spectroscopy, it can be deduced that (a) and (b) represent positively and negatively charged defects in BN, respectively.

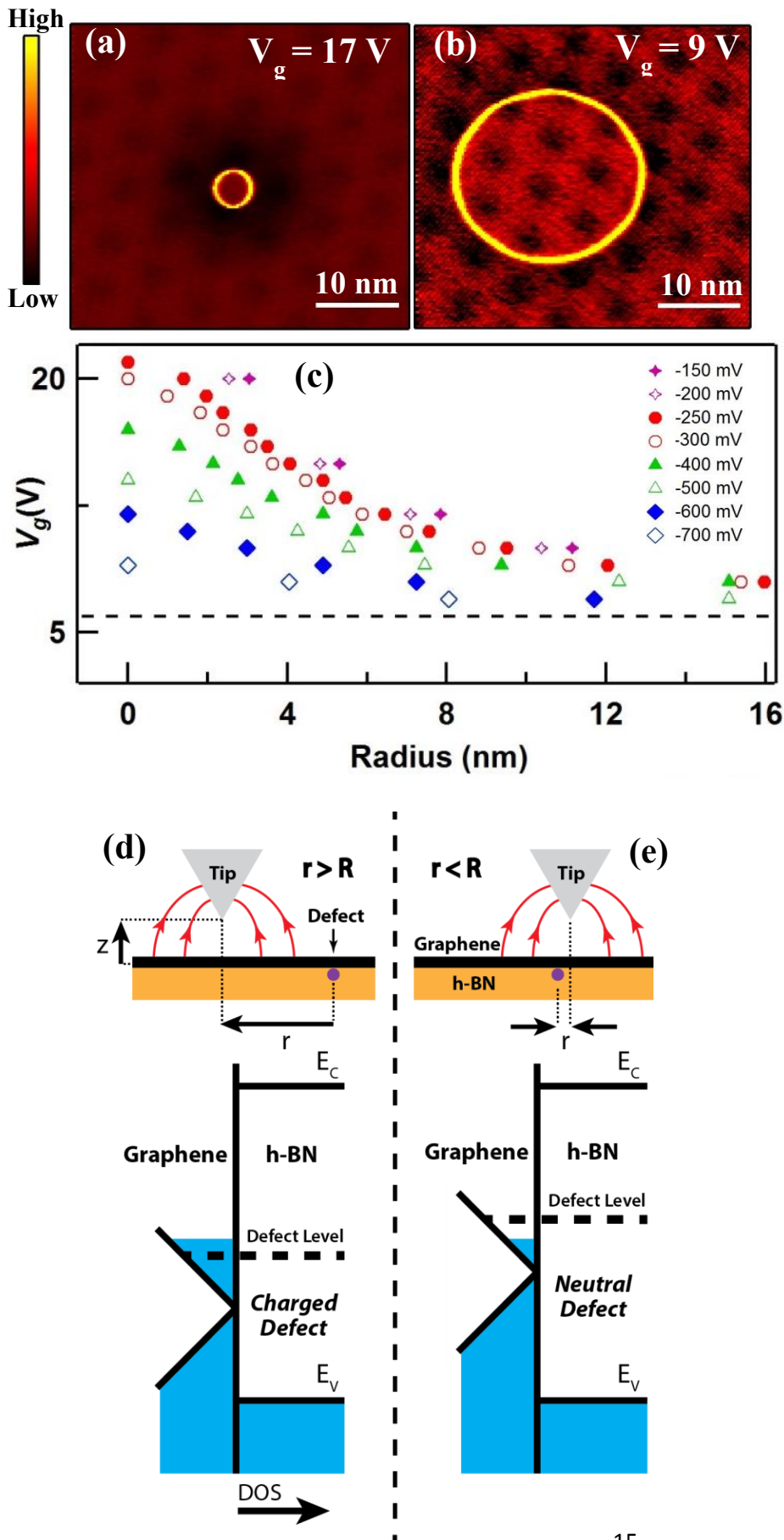

**Figure 3: d$I$/d$V$ maps of defect ring enable energy level characterization.** (a-b) d$I$/d$V$ maps ($I$ = 0.4 nA, $V_s$ = -0.3 V) of the same ring defect at back-gate voltages $V_g$ = 17 V and $V_g$ = 9 V, respectively. (c) Ring radius $R$ for different $V_s$ (denoted by distinct symbols) and $V_g$. The ring radius was extracted through d$I$/d$V$ maps all taken at the same location as (a-b). (d-e) Schematic model (energies not to scale) for ring formation due to charge transfer between graphene and a defect in the top layer of BN. When the distance $r$ between the tip and the defect is larger than $R$, the defect level is filled and negatively charged. When $r < R$, the local gating effect of the tip lowers the local electron density such that the Fermi level is below the defect level, neutralizing the defect. For negative tip potentials and n-doped graphene, $R$ increases as reduced $V_g$ shifts the unperturbed defect level closer to the Fermi energy.

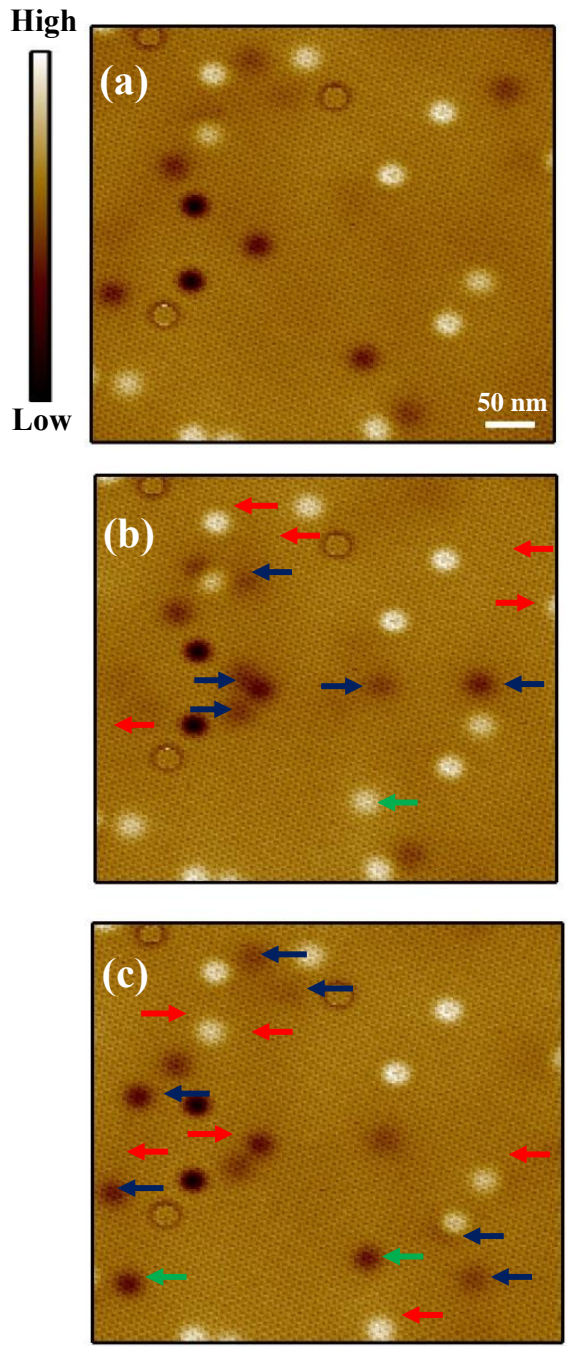

**Figure 4: Manipulating defects in BN with an STM tip.** Tip pulses having $V_s$ = 5 V are used to toggle the charge states of the dot defects. (a) d$I$/d$V$ map ($I$ = 0.4 nA, $V_s$ = -0.25 V) of graphene/BN reveals various dots and rings. (b) d$I$/d$V$ map of same region after a tip pulse is applied at the center of region in (a). (c) d$I$/d$V$ map of same region after another tip pulse. Red arrows mark the disappearance of dots relative to the previous image, blue arrows mark the appearance of dots, and green arrows mark dot defects that have changed the sign of their charge.